\documentclass[useAMS,usenatbib,usegraphicx]{mn2e}
\newcommand {\aplt} {{\raise-.5ex\hbox{$\buildrel<\over\sim$}}} 
 \title[A young hierarchical triple system harbouring a candidate debris disc]{A young hierarchical triple system harbouring a candidate debris disc\thanks{Herschel is an ESA space observatory with science instruments provided by European-led Principal Investigator consortia and with important participation from NASA.}\thanks{Based on observations collected at the German-Spanish Astronomical Center, Calar Alto, jointly operated by the Max-Planck-Institut fur Astronomie Heidelberg and the Instituto de Astrofisica de Andalucia (CSIC).} }
 \author[N.R.\ Deacon et al.]{N.R.\ Deacon\thanks{E-mail:deacon@mpia.de}$^1$, J.E.\ Schlieder$^1$, J.\ Olofsson$^1$, K.G. Johnston$^1$, Th. Henning$^1$\\ $^1$Max Planck Institute for Astronomy, Konigstuhl 17, Heidelberg, 69117, Germany\\
}
 \begin{document}
 \date{}
 \pagerange{\pageref{firstpage}--\pageref{lastpage}} \pubyear{2013}
 \maketitle
 \label{firstpage}
 \begin{abstract}
We report the detection of a wide young hierarchical triple system where the
primary has a candidate debris disc. The primary, TYC~5241-986-1~A, is a known Tycho star which we classify as a late-K star with
emission in the X-ray, near and far-UV and H$\alpha$ suggestive of youth. Its
proper motion, photometric distance (65--105~pc) and radial velocity lead us
to associate the system with the broadly defined Local Association of young
stars but not specifically with any young moving group. The presence of weak
lithium absorption and X-ray and calcium H and K emission support an age in
the 20 to $\sim$125~Myr range. The secondary is a pair of M4.5$\pm$0.5 dwarfs
with near and far UV and H$\alpha$ emission separated by approximately 1 arcsec
($\sim$65-105~AU projected separation) which lie of 145 arcsec (9200--15200~AU)
from the primary. The primary has a WISE 22~$\mu$m excess and follow-up
Herschel observations also detect an excess at 70~$\mu$m. The excess emissions are indicative of a 100--175~K debris disc. We also explore the possibility that this excess could be due to a 
coincident background galaxy and conclude that this is unlikely. Debris discs
are extremely rare around stars older than 15~Myr, hence if the excess is caused by a disc this is an extremely novel system.  
 \end{abstract}
 \begin{keywords} Astronomical data bases: Surveys -- optical\end{keywords}
\section{Introduction}
Since the late 1990s, several co-moving groups of young stars have been
identified in the Solar Neighbourhood \citep{Zuckerman2004}. These include the
$<$20~Myr old $\beta$~Pic moving group and TW Hydrae Association (TWA), the $\sim$30~Myr
Columba, Carina and Tucana-Horologium associations - which \cite{Torres2008}
group together as the Great Austral Young Association (GAYA) - and the AB~Dor
moving group which may share its origins with the Pleiades
\citep{Barenfeld2013}. These sit within a more general Local Association of
stars younger than or equal to approximately Pleiades age (125~Myr,
\citealt{Stauffer1998}). As well as providing targets for direct-imaging
studies of young extrasolar planets, these kinematic groups represent possible
remnants of low mass star formation events \citep{Mamajek2001}. As such they
are ideal targets to identify variations in the star formation process caused
by environmental dependence. Theoretical models (\citealt{Delgado-Donate2004},
\citealt{Kouwenhoven2010}, \citealt{Kaczmarek2011}, \citealt{Reipurth2012},) have suggested that dynamical interactions within a forming cluster can alter the population of binary stars. Evidence for such processing 
of binaries with separations $<$1,000~AU has recently been published by
\cite{King2012}. Hence any deviation in the wide ($>$1,000~AU) binary
population of such groups when compared to the field will lead to constraints
on the effect of formation environment dynamics on multiple systems. A number
of wide multiple systems have been identified in young moving groups
\citep{Torres2008} including a number of recently identified hierarchical
multiples \citep{Shkolnik2012}. However, there has been no systematic study of
the widest binaries in these associations.

Debris discs around stars are collections of dust radiating in the
mid-infrared to submillimetre. For $\sim$10~Myr old discs, this dust could be
the residuals of the steady state evolution of a protoplanetary disc
\citep{Wyatt2008}, the dust in older discs is thought to be accompanied by a
population of planetesimals and is to be replenished by collisions between
such objects \citep{Backman1993}. These planetesimals are the remnant of
planet formation processes much like our own Solar System's asteroid and
Kuiper belts. An observational connection between debris discs and planet
formation has been identified by \cite{Wyatt2012}. They find evidence that
stars with debris discs are more likely to host low-mass extrasolar planets.

While debris discs are commonly detected around stars of spectral types
A to early-K \citep{Su2006,Trilling2008,Moor2011}, their detection is substantially less
common in late-K or M dwarfs older than $\sim$20~Myr. Currently only two resolved debris discs
are known around M dwarfs, around the M1 $\beta$~Pic member AU~Mic
\citep{Kalas2004} and around the field age M5 GJ~581 \citep{Lestrade2012}. In
younger ($<$20~Myr) moving groups there are candidate debris discs with
mid-infrared excesses around late-K and M~stars such as $\beta$~Pic members AT~Mic \citep{Plavchan2009}
and GJ~182 \citep{Liu2004} and three TW~Hydrae members: TWA~3, TWA~7 \citep{Low2005}
and TWA~4 \citep{Skinner1992}. However, the number of observed late-type star (K5 or later) debris
discs drops off with age. \cite{Simon2012} find no WISE excess suggestive of a
disc in low mass members of Tucana-Horologium or AB~Doradus, while
\cite{Avenhaus2012} failed to identify WISE excesses caused by candidate discs
in any of the 103 field M dwarfs they studied. \cite{Fujiwara2013} studied 191
K dwarfs and 19 M dwarfs with photospheres detectable with AKARI 18$\mu$m data
and succeeded only in recovering TWA~4. However \cite{Heng2013} suggest that
is is due to the WISE satellite probing relatively small radii around
late-type stars at which dust is unlikely to survive for very
long. \cite{Currie2008} use Spitzer observations of the h and $\chi$ Persei
clusters to claim that 24~$\mu$m emission from debris disks peaks around
10--15~Myr before falling off.At longer wavelengths, \cite{Donaldson2012}
studied 6 K and M members of Tucana-Horologium with Herschel and found no
excesses. \cite{ps_smith2006} find a 70$\mu$m excess around the M2 dwarf
HD~95650 which \cite{King2003} list as kinematic and possible photometric
member of the $\sim$500~Myr old Ursa Majoris Moving Group as well as the K5
dwarf BD~+21$^{\circ}$~2486. Recently \cite{Eiroa2013} identified a cold,
resolved disk around the late-K star HIP~49908. In the $\sim$40~Myr old
cluster NGC~2547, \cite{Forbrich2008} find eleven M dwarfs with 24$\mu$m
excess. \cite{Lestrade2009} use these observations to derive a disc fraction
of 4.9$\pm$1.8\%, however \cite{Forbrich2008} state that their sample is not
complete for M dwarfs and that it is possible that their debris disk fraction
may be higher than for higher mass stars. \cite{Lestrade2006} find a cold
debris disc in the submillimeter around the $\sim$200~Myr star GJ~842.2. This
combined with their observations of their entire sample lead them to measure
cold disc fractions of 5.3$^{+10.5}_{-5.0}$\% for M dwarfs in the 20--200~Myr
age range and $<$10\% for older field objects. In summary, candidate debris
discs around late~K and M~dwarfs are rare after 20~Myr. However it is unclear if this is a true dearth or due to biases in the current searches. \cite{Lestrade2009} outline a number of possible mechanisms as to why M~dwarf discs are likely to disperse faster or be less observable than those around higher mass stars.
  \begin{figure*}
 \setlength{\unitlength}{1mm}
 \begin{picture}(100,120)
 \includegraphics{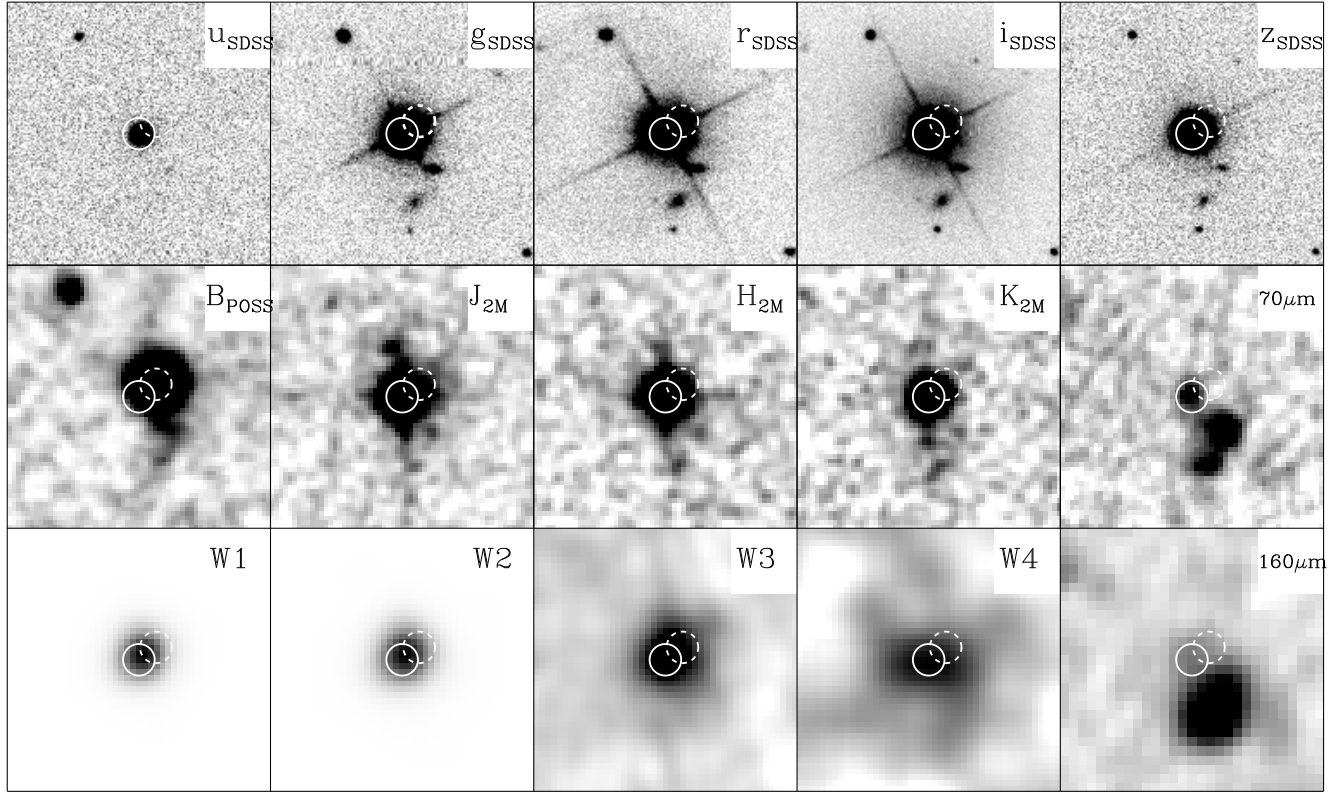}
 \end{picture}
 \caption[]{Multiband images of TYC~5241-986-1~A. The image size is one
   arcminute with the solid circle centred on the 2013.0 position and the
   dashed circle on the POSS-I $B$ position. Note the clear detection in the
   Herschel 70$\mu$m band. While this is slightly offset from the position of
   the star, the positional difference of 2.6 arcseconds is well comparable to
   the Herschel astrometric accuracy of 2.5 arcseconds. Note also the two faint spiral galaxies in the SDSS images, these have strong emission in the Herschel 70$\mu$m and 160$\mu$m bands. See Section~\ref{disc_galaxy} for more discussion on the possibility that these are part of a background galaxy cluster.}
 \label{primary_images}
 \end{figure*}
   \begin{figure*}
 \setlength{\unitlength}{1mm}
 \begin{picture}(100,120)
 \includegraphics{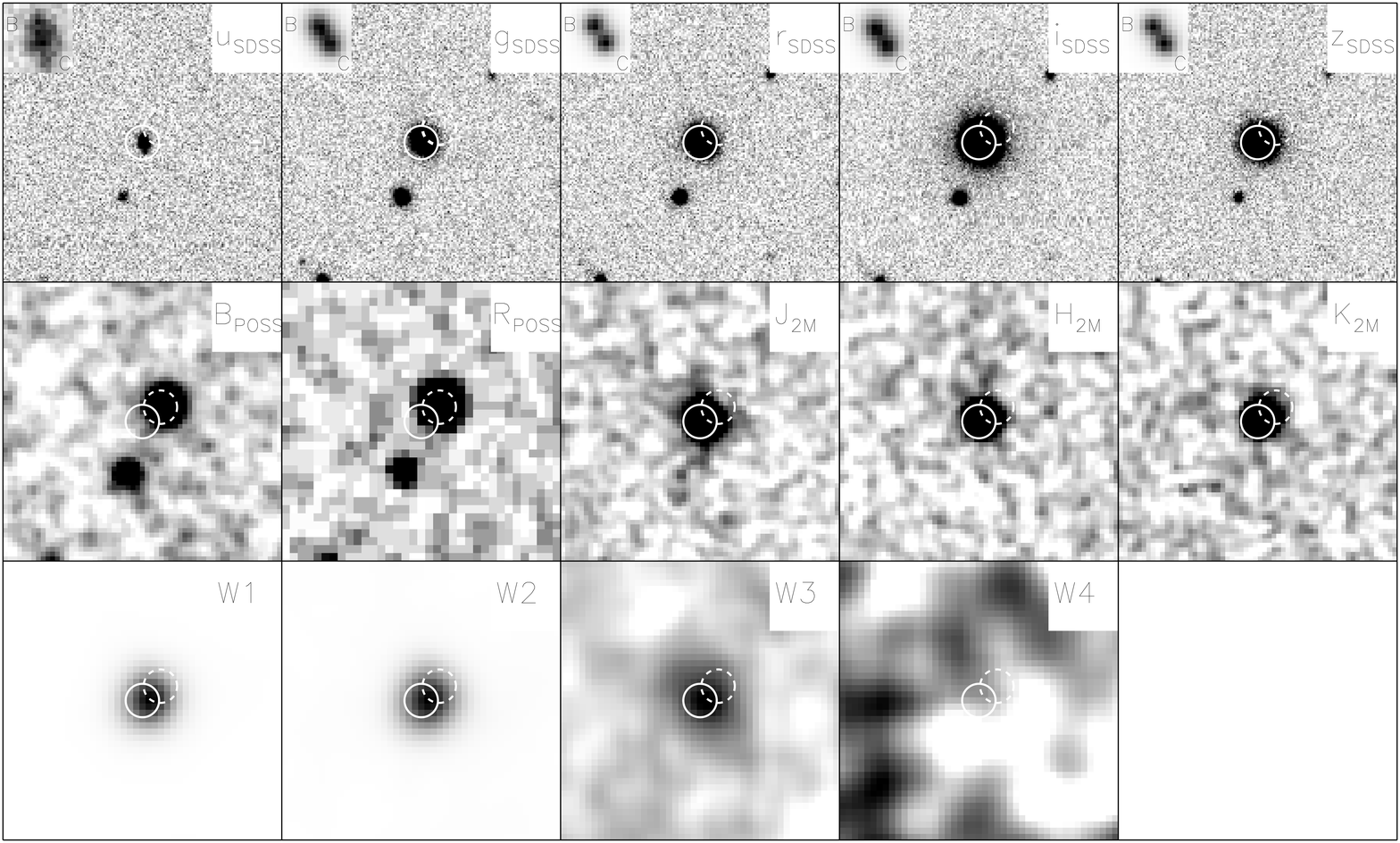}
 \end{picture}
 \caption[]{Multiband images of TYC~5241-986-1~B/C. The image size is one arcminute with the inset images 5 arcseconds across. In the SDSS inset images it can clearly be seen that the object is a binary with a separation of 1-2 arcseconds. Comparing with the POSS-I images from the 1950s it is clear that the two objects have moved with a common proper motion as there is no non-moving background star situated at the 2013.0 position (solid circle). The dotted circle marks the POSS-I $B$ position. Note also that the two components of this object appear to have near, equal flux throughout the SDSS bands.}
 \label{secondary_images}
 \end{figure*}
\cite{Schlieder2012} identified a list of Northern hemisphere candidate
members of the $\beta$~Pic and AB~Dor moving groups based on their proper
motions, photometry and UV and X-ray emission. We took a companion list of
candidate Southern hemisphere members of moving groups (Schlieder in prep.) and searched for companion objects in the SuperCOSMOS sky survey. In this paper, we report the discovery of a wide multiple system which has one component that harbours a possible debris disc.
\section{Identification of the system}
The input list in our search for wide companions consisted of 191 candidate
Southern hemisphere moving group members. These were selected using the
  same selection criteria as \cite{Schlieder2012} with objects initially
identified as 1: having a proper motion vector aligned to within ten
  degrees of the local projection of the space velocity of the moving group, 2:
  colour-magnitude diagram placement consistent with other moving group members
when using a kinematic distance estimate that assumes group membership, 3: having  a $(V-K_s) \geq 3.2$ (suggesting
that they are M dwarfs) and 4: having emission in UV
($\log(F_{NUV}/F_{K_s}) \geq –4.1$ or $\log(F_{FUV}/F_{K_s}) \geq –5.1$) or
X-ray ($\log(F_X/F_{K_S}) \geq –2.6$) indicative of youth\footnote{Note that these are only candidate members of a particular moving group. Follow-up observations are required for each to determine the full 3-D kinematics and to measure youth indicators before their membership can be confirmed.}. The target primary stars come from a Southern extension to
\cite{Lepine2005} (L\'{e}pine private communication). The stars are candidate
members of TWA, $\beta$ Pic, Tucana-Horologium and AB Dor. However as many
young associations have similar kinematics and ages, these objects may also be
candidate members of other moving groups. We ran an initial search on the
SuperCOSMOS Science Archive within 5 arcminutes of each candidate star and
identified objects with proper motion measurements more significant than 5$\sigma$ and which met the GALEX \citep{Martin2005} far and/or near-UV activity cuts set out by \cite{Schlieder2012}. We identified 2652 objects within 5 arc minutes of our targets which passed our proper motion significance cuts, 220 of these passed our UV cut. We then plotted the significance of the proper motion difference\footnote{The quadrature sum of the proper motion difference in each axis divided by the total error in that axis.} between the SuperCOSMOS proper motions of both objects against their projected separations. This is shown in Figure~\ref{TYC_shifted}. Note two objects stand out, one is TYC 7443-1102-1~B, a known wide companion to a $\beta$ Pic member from \cite{Lepine2009}. The other obvious outlier is TYC~5241-986-1, a 
star in the Tycho catalogue. It appears to be a 145 arcsecond companion to
PM~I22595-0704, a candidate member of the Tucana-Horologium moving group based on it's proper motion and colour-magnitude diagram placement (Schlieder et al. in prep.). We
found this star had emission suggestive of activity (and hence youth) in the
X-ray region from ROSAT \citep{Voges2000} and in the near and far-UV from
GALEX \citep{Martin2005}. TYC~5241-986-1 was not in our input list of
candidate moving group members as although it has sufficient X-ray and UV
emission to pass those cuts, it had a $V-K_s$ colour that was bluer than the
cut defined by \cite{Schlieder2012}. This $V-K_s \geq 3.2$ cut was due to their
  search being specifically for M dwarf members, our search includes no such
  cut and hence can include earlier type companions. Images of TYC~5241-986-1 in multiple filters are shown in Figure~\ref{primary_images} with similar plots for PM~I22595-0704 in Figure~\ref{secondary_images}. Inspecting Sloan Digital Sky Survey \citep{SDSS_DR9} images of PM~I22595-0704 revealed that it is itself a close binary system with a flux ratio close to unity. For more details on both components, see Table~\ref{objects}. As TYC~5241-986-1 is a previously catalogued object, we will refer to the entire system as TYC~5241-986-1~A and TYC~5241-986-1~B/C.

After extracting the WISE \citep{Wright2010,WISE2012} photometry for TYC~5241-986-1~A, we noted it had a particularly striking $W3-W4$ colour of 1.64 magnitudes. This combined with it being colour neutral in colours involving $W1$, $W2$ and $W3$ makes it similar to the debris disc star HD~191089 \citep{Mannings1998}. This along with its possible membership of a young moving group with an age of 30~Myr~\citep{Zuckerman2004} makes it an interesting target for further characterisation.

\begin{table*}
 \begin{minipage}{170mm}
  \caption{Details of the photometry and astrometry for the A and integrated
    B/C components of the TYC~5241-986-1 system. Citation key:$^a$
    \protect\cite{Skrutskie2006},$^b$ \protect\cite{Hambly2001}, $^c$
    \protect\cite{Hog2000},  $^d$
    S. L\'{e}pine (private communication) calculated using photographic plate magnitudes and
    the method of \protect\cite{Lepine2005}, $^e$ \protect\cite{Wright2010,WISE2012}, $^f$ \protect\cite{Voges2000}, $^g$ \protect\cite{Martin2005}, $^h$ this work.}
\label{objects}
\begin{center}
 \footnotesize
  \begin{tabular}{lcc}
  \hline
&TYC~5241-986-1~A&TYC~5241-986-1~B/C\\
&&(integrated)\\
  \hline
Position (J2000)&22$^h$59${'}$34.99${\arcsec}$ $-$07$^{\circ}$02${'}$22.7${\arcsec}$
$^a$&22$^h$59${'}$34.84${\arcsec}$ $-$07$^{\circ}$04${'}$46.7${\arcsec}$ $^a$\\
Epoch&1998.841$^a$&1998.841$^a$\\
$\mu_{\alpha}\cos\delta$(mas/yr)&69$\pm$11$^b$&70$\pm$9$^b$\\
&67.0$\pm$3.8$^c$&\\
$\mu_{\delta}$(mas/yr)&$-$57$\pm$10$^b$&$-$51$\pm$9$^b$\\
&$-$53.7$\pm$3.5$^c$&\\
$B_J$ (mag)&12.33$^b$&16.17$^b$\\
$B$ (mag)&12.78$\pm$0.29$^c$&\\
$V$ (mag)&11.59$\pm$0.14$^c$&15.4$^d$\\
$R$ (mag)&10.48$^b$&13.91$^b$\\
$I_N$ (mag)&9.53$^b$&11.68$^b$\\
$J$ (mag)&9.47$\pm$0.03$^a$&11.01$\pm$0.04$^a$\\
$H$ (mag)&8.77$\pm$0.05$^a$&10.38$\pm$0.03$^a$\\
$K_s$ (mag)&8.67$\pm$0.02$^a$&10.13$\pm$0.03$^a$\\
$W1$ (mag)&8.56$\pm$0.03$^e$&9.84$\pm$0.04$^e$\\
$W2$ (mag)&8.58$\pm$0.02$^e$&9.65$\pm$0.02$^e$\\
$W3$ (mag)&8.50$\pm$0.05$^e$&9.36$\pm$0.10$^e$\\
$W4$ (mag)&7.36$\pm$0.18$^e$&$>$8.4$^e$\\
Herschel/PACS 70 $\mu$m (mJy)&$7.1 \pm 1.8$&\\
log(f$_X$/f$_{K_s}$)&-1.93$^f$&undetected $^f$\\
log(f$_{NUV}$/f$_{K_s}$)&-3.45$^{a,g}$&-3.90$^{a,g}$\\
log(f$_{FUV}$/f$_{K_s}$)&-4.49$^{a,g}$&-4.03$^{a,g}$\\
Spectral Type&late-K$^h$&M4.5$\pm$0.5$^h$\\
  \hline

\normalsize

\end{tabular}
\end{center}
\end{minipage}
\end{table*}
\subsection{Companionship}
\begin{figure}
 \setlength{\unitlength}{1mm}
 \begin{picture}(100,60)
 \includegraphics{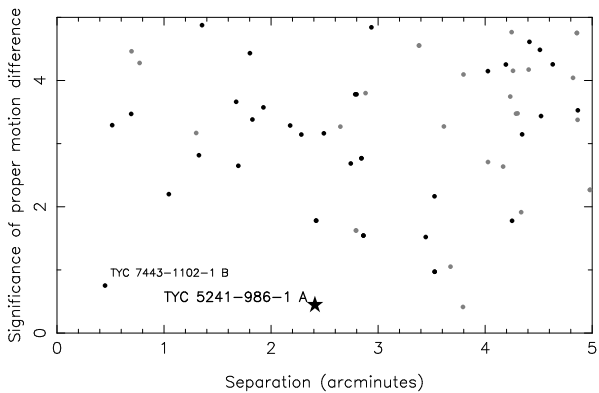}
 \end{picture}
 \caption[]{TYC~5241-986-1~A shown with other stars selected by the same
   proper motion and UV excess cuts (black points). The y-axis is the
   quadrature sum of the proper motion differences in each axis divided by the
   error on the proper motion measurement in that axis. The grey points show objects paired together using the offset method of \cite{Lepine2007}, these should be random pairings. Its appears that TYC~5241-986-1~A lies outside the coincident distribution. Note also our recovery of TYC 7443-1102-1~B \citep{Lepine2009} as a wide companion.}
 \label{TYC_shifted}
 \end{figure}

  There are several checks we can apply to infer if these objects are a likely pair. Firstly \cite{Dupuy2012} asserted that objects with $\delta \mu/ \mu > 0.2$ were unlikely to be a true pair. This system has a fractional proper motion difference of 0.07. \cite{Lepine2007} derived several metrics to judge if a pair was a true pair or a pairing with a background star with coincident proper motion. They produced a purely coincident population by offsetting the positions of their input list by angles of a few degrees to generate pairings with background stars. We applied the same method by shifting the R.A. coordinates of our input list by two degrees and then following through the proper motion significance and UV cuts we applied to identify the star. The results are shown as grey dots in Figure~\ref{TYC_shifted}. Clearly our system lies well outside the coincident population. \cite{Lepine2007} also provide an inequality comparing the product of the proper motion and angular separation with the total 
proper motion of the system ($\Delta\mu \times separation < (\mu/0.15)^{3.8}$ ). This is based on the distribution of proper motions of all stars in their catalogue and is equal when an object has a 50\% chance of being a coincident pairing. Hence if an object passes the inequality, it is likely a companion. If we apply this to our system we find that it fails their condition with values of 0.88 and 0.138 for the two sides of the inequality. However the right-hand side of the \cite{Lepine2007} inequality is derived from pairings of all stars in their proper motion catalogue. We know that our two stars have UV emission suggestive of youth and hence are only drawn from a young subset of stars in the sky. Hence we must correct the right-hand side of the inequality to take this lower surface density of coincident pairings with young stars in to account. Only 220 of the 2652 objects which passed our proper motion significance cut had sufficient emission to pass our UV cut. If we adjust the value of $(\mu/0.15)^{3.8}$ by a factor of 2652/220, then our system satisfies the \cite{Lepine2007} inequality.
\section{Characterisation of the components}
\subsection{TYC~5241-986-1~A}
\begin{figure}
 \setlength{\unitlength}{1mm}
 \begin{picture}(100,80)
 \includegraphics{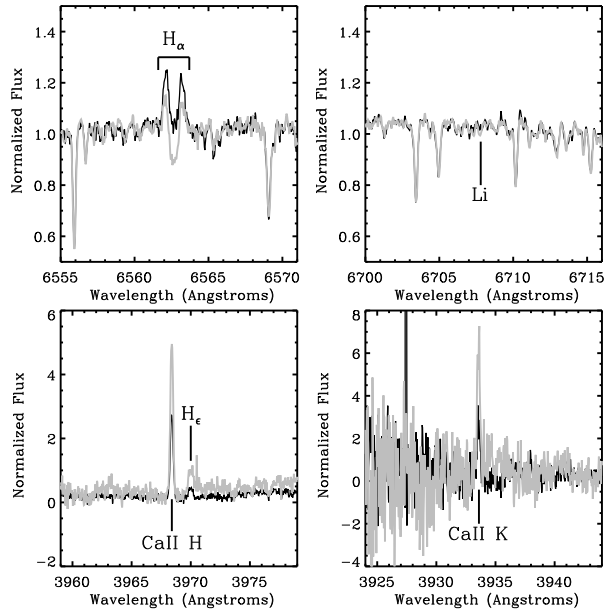}
 \end{picture}
 \caption[]{Portions of our FEROS spectrum for TYC~5241-986-1~A. the black
   line represents the spectrum taken on the 6th of July 2012~UT and the grey
   line the spectrum from the 28th of August 2012~UT. The star has emission in
   H$\alpha$, H$\epsilon$ and calcium H and K (all shown). The calcium K line
   lies at the blue end of a FEROS order and is thus in a noisy region of the spectrum. Additionally there
   is emission in the core of the three lines of the Ca~II infrared
   triplet. We interpret the H$\alpha$ profile as emission with
   self-absorption. The 6707.8~\AA~lithium feature is extremely weak having
   an equivalent width of $\sim$10~m~\AA. In the first epoch spectrum this
   appears to be contaminated by a noise spike but is much clearer in the
   second epoch. Note the H$\alpha$ self-absorption and H$\epsilon$ and
   calcium H and K emission are all stronger in the second epoch.}
 \label{TYC_A_spectrum}
 \end{figure}

We obtained $R$=48000 spectroscopic observations of TYC~5241-986-1~A using
FEROS \citep{Kaufer1999} on the MPG/ESO 2.2~m telescope in La~Silla, Chile on
the 6th of July 2012 (UT). Three integrations of 500~s were used. The data
were reduced using standard ESO-MIDAS FEROS packages and the normalised
spectrum was cross-correlated over a number of orders with a suite of
late-type templates from \cite{Prato2002} which have subsequently been observed
using FEROS with the same setup. This produced an RV measurement of
-6.7$\pm$0.1~km\,s$^{-1}$ fitting K5 and M0 templates well. However when we examined
the star's H$\alpha$ profile, we were surprised to find two peaks above the
continuum (see Figure~\ref{TYC_A_spectrum}). Other prominent features in the
spectrum did not appear to be double lines. To help understand the cause of
this we re-observed the target using the same set-up on the 18th of August
2012 (UT). This time the H$\alpha$ profile appears as two emission features
either side of the expected wavelength with absorption in between. Such
self-absorption of H$\alpha$ emission is commonly seen in active
M~dwarfs~\citep{Worden1981,Houdebine1995,Mauas2000}.  The radial velocity 
measurement for this second spectrum was -6.7$\pm$0.1~km\,s$^{-1}$. In addition to
H$\alpha$, emission was also seen in H$\epsilon$, calcium H and K and the
Ca~II infrared triplet in both spectra. These indicate that this is an extremely active
star. It is notable that the depth of the H$\alpha$
self-absorption and the height of the emission in other lines appear to be
stronger in the second spectrum. Figure~\ref{TYC_A_spectrum} shows cutouts for some of the regions with
emission lines as well as the 6707.8~\AA~lithium absorption feature. This is
weak, measuring only $\sim$8~m\AA~in equivalent width averaged over our two spectra. As our spectrum is of such high resolution, there are hardly and regions of continuum to measure the noise from. Hence we use our two independent measurements from our two spectra finding a scatter on the measurements of approximately 4~m\AA. \cite{Mentuch2008}
show lithium measurements for some late~K and early~M dwarfs in $\beta$~Pic of
a few tenths of an Angstrom. Tucana-Horologium shows a mixture of values from
these types of stars, from a few tenths of an Angstrom to some measurements
which are in the 40--50~m\AA~range, while AB~Dor members in this spectral
range show lithium equivalent widths mostly below 50~m\AA~and often
consistent with zero. \cite{Torres2008} indicate that a star of this colour
($V_c-I_c=1.9$ magnitudes) is close to the lithium depletion boundary for the Pleiades. Given that we have a very weak lithium 
detection we find it unlikely that our object is substantially older than this
cluster ($\sim$125~Myr; \citealt{Stauffer1998}). We can also infer our object
is probably older than $\beta$~Pic and may be older than
Tucana-Horologium. While lithium depletion is not a simple monotonic clock for
stellar ages, the weak lithium absorption in this star indicates that it is more likely to be older than 20~Myr than substantially younger. We estimated the flux in the calcium H and K lines by measuring the equivalent width and converting to flux using the relations of \cite{Hall1996}. This yielded values of $\log R'_{HK}=$-3.63 and -3.8 for our first and second spectra respectively. This is consistent with stars of a similar spectral type in the Pleiades \citep{Soderblom1993,King2003}, but inconistent with members of Ursa Majoris \citep{Lopez2010}. The latter paper also studied a number of proposed members of AB~Dor, these have values comparable with TYC~5241-986-1~A. \cite{Mamajek2008} find that $\log R'_{HK}$ saturates at approximately 
the value we find for TYC~5241-986-1~A. Hence we cannot use this to set a
lower age boundary. From these constraints, we can say that the TYC~5241-986-1
system is likely older than 20~Myr and younger than the 500~Myr old Ursa
Majoris group. Its activity and lithium measurements match the members of
AB~Dor and the Pleiades well, indicating that it may be of similar age ($\sim$125~Myr) and unlikely to be significantly older.

We estimated the rotational velocity TYC~5241-986-1~A by comparing its
spectrum to artificially broadened versions of our template spectra. This was
done in steps of 2~km\,s$^{-1}$ with a $v \sin i$ of 8~km\,s$^{-1}$ giving the
best correlation. We then calculated the maximum rotational period at 20 and 125~Myr by
combining the rotational velocity with radii calculated from the
\cite{Baraffe1998} models. This resulted in maximum rotational periods of
$\sim$6~days for an age of 30~Myr and $\sim$4~days for
125~Myr. \cite{Eyer2006} identified TYC~5241-986-1~A as a photometric variable
with a period of 8 days. As this is a measured photometric period we prefer
this over our period determined from rotational velocity. Comparing with Figure~9 from
\cite{Mamajek2008} shows that for both ages, the star would fall on the slower
rotator sequence for the Pleiades with a period too short to
be consistent with the
Hyades.  This is another indicator that TYC~5241-986-1~A is unlikely to be
significantly older than the Pleiades.

These constraints in addition to TYC~5241-986-1~A having significant UV excess (suggested by \citealt{Shkolnik2011} as an indicator of an object being younger than 300~Myr) lead us to place an age range on this object of 20 to $\sim$125~Myr.

 As we do not have a flux calibrated single order spectrum of
 TYC~5241-986-1~A, we used photometric methods to estimate the spectral
 type. Fitting our $B$, $V$, $J$, $H$, $K_s$, $W1$, $W2$ and $W3$ photometry to the model photospheres of \cite{Castelli2004}, we find a
 best fitting temperature of 4200~K. We also compared the $V-K$ colour of the
 object to empirical relations from \cite{Kenyon1995}, finding
 4380$\pm$150~K. These two effective temperature values convert to spectral
 types of K6 and K5$\pm$1 respectively (using \citealt{Kenyon1995}). To
 estimate the distance to TYC~5241-986-1~A, we compared the 2MASS photometry
 for the primary with models for populations with ages of 20 and 130~Myr from
 \cite{Baraffe1998}. For the 20~Myr age and our two spectral type values, we
 find a distance range 85--105~pc. By the time our object has reached
 125~Myr it is close to being on the main sequence. Hence for the 130~Myr
 models we derive distances of approximately 65~pc. As these are photometric
 distance estimates we note that this range is likely only accurate to approximately 20\%.

We estimated the X-ray luminosity of the star by combining its ROSAT
\citep{Voges2000} count rate and hardness ratio using the equation of
\cite{Schmitt1995}. This yielded a value of
$L_x=4.13\pm1.39\times10^{29}$~ergs\,s$^{-1}$ for our approximate 20~Myr distance of
 95~pc and $L_x=1.94\pm0.72\times10^{29}$~ergs/s for 65~pc (our $\sim$130~Myr
distance). We then took the bolometric magnitude from the \cite{Baraffe1998}
models and used this to calculate $R_x$. For the 20~Myr case we derived $\log
R_x=-3.4$ and -3.3 for the $\sim$125~Myr. Again this is in the range of
activity values \cite{Mamajek2008} find to be too saturated to be a useful age
indicator. All we can say from this value is that it is consistent with our previous age range of 20 to $\sim$125~Myr.
  
\subsection{TYC~5241-986-1~B/C}
Spectroscopic observations of TYC~5241-986-1~B/C were obtained with the CAFOS instrument on the Calar Alto 2.2m telescope on the 8th of December 2012. Two 300s exposures were obtained using the R-100 grating with the slit orientated to maximise the distance between the components in the resulting spectrum. An observation of the standard star BD+25$^{\circ}$4655 was also obtained at similar airmass.

\begin{figure}
 \setlength{\unitlength}{1mm}
 \begin{picture}(100,80)
 \includegraphics{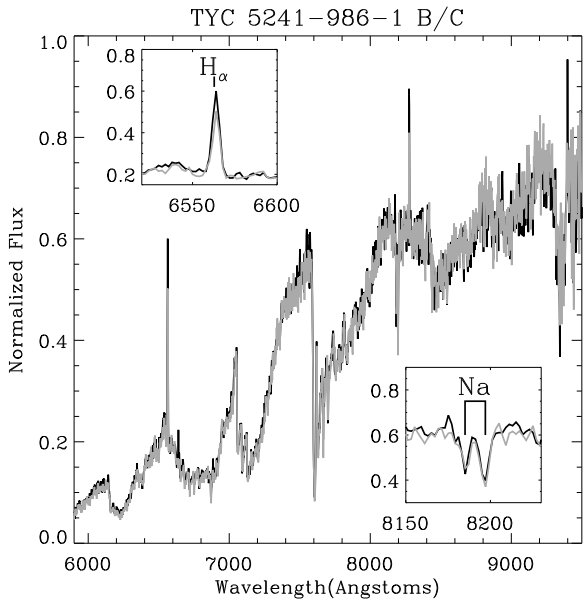}
 \end{picture}
 \caption[]{A plot of the spectra of both component B (black) and C (grey) of
   the TYC~5241-986-1 system. The insets show the H$_{\alpha}$ emission for
   both components and Sodium 8200\AA~doublet. We type both components as
   active M4.5 dwarfs with an error of half a subtype (see text for
   details). The sodium doublet has equivalent widths of $-3.5\pm0.2$~\AA~and
   $3.3\pm0.3$~\AA~for components B and C respectively. Comparing with measurements in \cite{Schlieder2012a} we conclude that the system is younger and has a lower gravity than typical field M dwarfs.}
 \label{sec_spec}
 \end{figure}

The resulting spectrum was heavily blended with a separation between the peak
of each component of approximately 2 pixels. Firstly, the images were bias
subtracted and flat fielded using standard IRAF routines. Then we fitted a low
order polynomial to each wavelength (excluding the two spectra) and subtracted
this to remove background sky emission. The same procedure was followed for
the standard star. We then employed the technique of \cite{Hynes2002} to
separate the blended spectra. The standard was used to define a model spectral
profile. For each wavelength (i.e. row on the chip parallel to the slit), we
fitted a Voigt function to the standard spectrum. This allowed us to derive
how the parameters of the Voigt profile which describes the spectrum in the
direction parallel to the slit vary with wavelength. The distributions of
these parameters against wavelength were then smoothed by fitting a low order polynomial to each. Then for each wavelength of the target star observation, we defined a model blended spectrum to be 
the sum of two Voigt profiles each with their own normalisation and positional offset parallel to the slit. We then ran a Markov Chain Monte Carlo analysis to find the best fit model parameters for both components at each wavelength. The spectrum for each component at each wavelength is the integral of the fitted Voigt profile for that component at that wavelength. The standard star was extracted using a similar technique but with only the normalisation as a free parameter. The spectra were then wavelength calibrated using arc spectra and IRAF routines. Finally flux calibration was carried out by comparing the observed continuum flux of the standard with the measured values from \cite{Oke1990}. A low order polynomial was fitted to the derived sensitivity function and this was used to correct the observations of the target. The final spectra of both components averaged over the two observations is shown in Figure~\ref{sec_spec}. It has an approximate noise of 3 counts based on regions of the pseudocontinuum.

We measured the spectral indices defined by \cite{Lepine2003} and used their
index to spectral-type relations to derive the spectral type of the B and C
components. Table~\ref{indices} shows the spectral index measurements along
with the spectral type associated with each of these measurements. Based on
these we derive a spectral type of M4.5$\pm$0.5 for both components. We also
note that H$_{\alpha}$ is in emission for both components with equivalent
widths of $-9.9\pm0.3$~\AA~and $-7.5\pm0.2$ \AA~for components B and C
respectively. However given the spectral type of both components, this does
not set a strong constraint on their age. \cite{West2008} list a 1$\sigma$
upper bound of 7.5Gyr for an M5 star (the upper bound of our spectral type
range). The equivalent widths of the gravity sensitive sodium 8200~\AA~doublet are $-3.5\pm0.2$~\AA~and $3.3\pm0.3$~\AA. These values are comparable
to those of two proposed $\beta$~Pic members of similar spectral type observed by
\cite{Schlieder2012a}. While this spectral feature cannot be used 
to set a strong age constraint, we note that these values make it likely that
TYC~5241-986-1~B/C is younger than the typical population of field M
dwarfs. In common with the primary these stars show UV excess, making them
likely to be younger than 300~Myr. We followed a similar process to the
primary when estimating the photometric distance to the secondary. One
deviation was that we adjusted the 2MASS photometry by 0.75 magnitudes to take
into account that this approximately equal-flux binary is unresolved in this
dataset. From this we estimate the photometric distance to be in the range
65--90~pc. Taking into account the likely $\sim$20\% error on this photometric
distance, this is in agreement with the values of approximately 65~pc
calculated for the primary at an age of 125~Myr and the 20~Myr distance value
of 85--105~pc. The similarity in photometric distances supports
TYC~5241-986-1~B/C's companionship with the primary. 

\subsection{The architecture of the system}

Comparing our approximate effective temperatures for all three components with
the \cite{Baraffe1998} models we find masses of $\sim$0.7~M$_{\odot}$ for the
primary and 0.15--0.175~M$_{\odot}$ for each of the secondary components. The
two lower mass components of the system are separated by 145'' from the
primary, equating to a distance in the 9200--15200~AU range. Comparing the
total mass of the system and projected separation with the field binaries
plotted in Figure~15 from \cite{Close2003}, we find that this system is
more loosely bound than any known system of similar mass within 25~pc. This
indicates that this system may not survive to field age. 

The fact that this is a hierarchical triple system is also
interesting. \cite{Dhital2013} and references therein
(i.e. \citealt{Fischer1992}, \citealt{Reid1997})find that wide binaries
have a far higher multiplicity fraction than field stars. This is taken as an
indication that dynamical ejection of the low-mass pair played a role in
forming the system.

\begin{table*}
 \begin{minipage}{170mm}
  \caption{The derived spectral indices for the B and C components of the system. The indices used here are defined by \protect\cite{Lepine2003} and the spectral classifications in brackets are derived from the spectral type relations for each index. Note we did not use the TiO6 feature due to significant telluric contamination.}
\label{indices}
\footnotesize
  \begin{tabular}{lccccccc}
  \hline
Component&CaH2&CaH3&TiO5&TiO7&VO1&VO2&ColourM\\
  \hline
B&0.378 (M4.0)&0.668 (M3.8)&0.365 (M4.2)&0.851 (M4.3)&0.881 (M5.3)&0.736 (M4.7)&2.793 (M4.9)\\
C&0.362 (M4.3)&0.632 (M4.4)&0.347 (M4.4)&0.841 (M4.4)&0.894 (M4.9)&0.711 (M4.9)&2.900 (M5.0)\\
  \hline

\normalsize
\end{tabular}
\end{minipage}
\end{table*}
\subsection{Moving group membership}
 \begin{figure*}
 \setlength{\unitlength}{1mm}
 \begin{picture}(100,120)
 \includegraphics{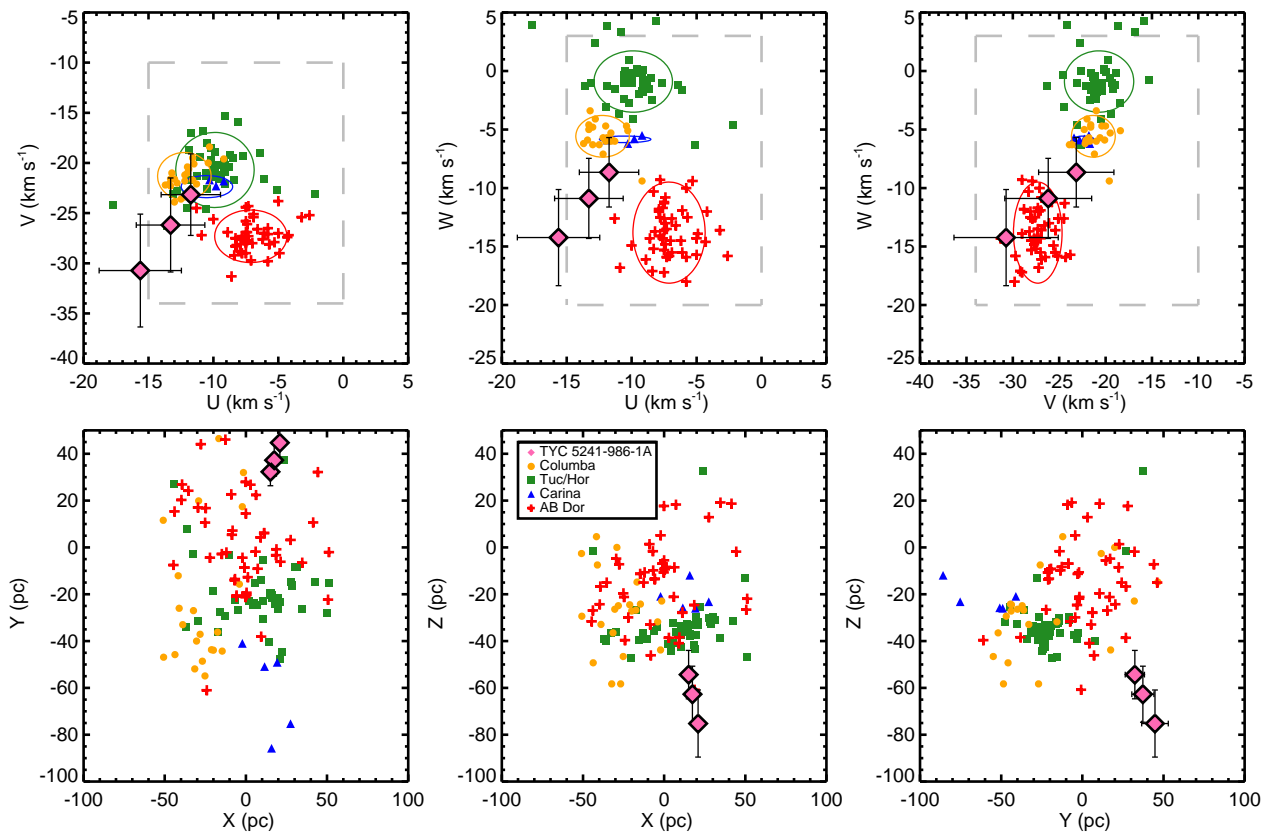}
 \end{picture}
 \caption[]{U,V,W,X,Y, and Z projections for TYC~5241-986-1~A compared to several known young kinematic
associations. TYC~5241-986-1~A is represented as a large diamond with associated error bars.
The three data points, with smallest to largest errors, are for photometric
distances/ages of 65~pc/130 Myr, 75~pc/80 Myr, 90~pc/20 Myr. The known group
distribution symbol designations are given in the legend and the solid ellipses with
like colors represent the 2$\sigma$ errors on their average velocties. The dashed gray
box represents the kinematic space occupied by nearby, young stars known as the Local
Association. The top panels show that TYC~5241-986-1~A is only consistent with the 2$\sigma$
error ellipse of the Columba and Carina associations in all three velocity projections, although
at a photometric distance (65~pc) that corresponds to an age that is inconsistent
with these associations. The bottom panels show that the galactic distances of TYC~5241-986-1~A
are marginally consistent with the distributions of several associations. This is
not a strong constraint on membership. Without a parallax measurement, we cannot
suggest membership in any individual group and can only conclude that the TYC~5241-986-1~A
system has kinematics and age consistent with the Local Association.}
 \label{TYC_A_MG}
 \end{figure*}

Since TYC~5241-986-1~B/C was originally selected as a candidate of the Tucana/Horologium
association based on its proper motion and photometry, we investigated its membership
in several nearby kinematic associations. Given that we constrain the age of
the system to be $\sim$20 to $\sim$125 Myr, we compare to five kinematic associations with
comparable ages, the Tucana/Horologium, Columba, and Carina associations ($\sim$30 Myrs, \citealt{Torres2008}),
the AB Doradus moving group ($\sim$125 Myr
\citealt{Zuckerman2004,Torres2008,Barenfeld2013}), and the Local Association
($\aplt$125 Myr, \citealt{Zuckerman2004} and references therein). We followed the
relations in \cite{Johnson1987} to calculate the U,V,W galactic velocities
and X,Y, and Z galactic distances of TYC~5241-986-1~A from its published proper motion,
measured radial velocitiy and photometric distance\footnote{we define U and X
positive toward the Galactic center, V and Y positive in the direction of solar
motion around the Galaxy, and W and Z positive toward the north Galactic
pole.}. Figure~\ref{TYC_A_MG} shows the 6D kinematic distributions of the five previously mentioned
associations \citep{Malo2013,Zuckerman2004} and TYC~5241-986-1~A. The
top three panels are projections in U,V, and W velocity, the bottom three are
projections in X,Y, and Z distance. We represent TYC~5241-986-1~A as a large diamond symbol
with associated error bars. The known kinematic groups are represented by symbols
defined in the figure legend with ellipses of matching colour in the U,V, and W
projections defining the 2$\sigma$ error on the average velocities of the association
\citep{Malo2013}. The dashed gray line represents the kinematic space of the Local
Association of nearby, young stars \citep{Zuckerman2004}. TYC~5241-986-1~A is shown at
three different estimates of the photometric distance which are indicative of
different possible ages of the system, 65~pc - 130 Myr, 75~pc - 80 Myr, 90~pc - 20
Myr. In the top panels, the only groups the Galactic velocities of
TYC~5241-986-1~A are consistent with in all three projections are the Columba and Carina associations at a distance of ~65~pc and the
more general Local Association. However the 65~pc photometric distance corresponds to an age that is
not consistent with the proposed ages of Columba and Carina. The bottom panels show that the
Galactic distances of TYC~5241-986-1~A are possibly consistent with several of the
associations, but it lies on the periphery of the distributions in any case. As a
supplement to the comparisons in the figure, we calculate the probability that TYC~5241-986-1~A is a member of one of the associations using the Bayesian analysis tools of
\cite{Malo2013}. We found the highest membership probability over our range of
photometric distances is with the Columba association, although only ~30\% at 65~pc (a distance inconsistent with TYC~5241-986-1~A being of similar age to Columba).
Thus, without a measured parallax to better constrain the system's distance, we cannot
suggest any membership assignment to any of the individual associations and can only
conclude that the TYC~5241-986-1 system's kinematics and youth are consistent with other
nearby young stars in the solar neighbourhood that comprise the Local Association.

\section{Mid-infrared excess}
\subsection{Fidelity of the WISE data}
In order to check the quality of the WISE $W4$ excess, we performed a number
of checks. Firstly, we examined the WISE images to ensure that the object
appeared as a point-like source in all four bands (3.4, 4.6, 12, 22~$\mu$m). Next we checked the aperture magnitude in the $W4$ band. This was found to be 7.56$\pm$0.32, in good agreement with the PSF magnitude of 7.36$\pm$0.18. We then examined the individual measurements, we found that there were several anomalously bright detections in the catalogue for this source. As the source was flagged as having several measurements contaminated by scattered moonlight, we contacted the WISE helpdesk who informed us that the anomalous measurements had not been used in the final WISE coadd. Five images were used to produce the final $W4$ catalogue values for this star. In two of these it was detected at a significance of over 3$\sigma$, this makes a brief transient event unlikely as the cause for the excess. Additionally the source has a $\chi^2$ goodness of fit statistic of 1.06 for the $W4$ 
band profile fit, further supporting the fidelity of the detection. Hence we
assume that the excess in $W4$ is real and has an astrophysical cause.
\subsection{Herschel observations}
It is virtually impossible to characterise the excess based on
only a single datapoint. Additionally, while we have exhaustively endeavoured to confirm the reliability of the WISE
$W4$ excess, there is always the possibility of a spurious measurement for reasons we have not considered. Hence we obtained Herschel PACS 70~$\mu$m and 160~$\mu$m observations. 

We observed TYC~5241-986-1\,A with the PACS photometer \citep{Poglitsch2010} onboard Herschel \citep{Pilbratt2010} on the 31st of December 2012. The observing strategy consisted of two consecutive \textsc{Pacs} mini-scan maps (program ``OT1\_jolofsso\_1'', OBSIDs 1342257974 and 1342257975) in the blue filter (70\,$\mu$m). The observing duration was $2470$\,sec per mini-scan map. When observing with \textsc{Pacs} with the blue camera, data is also simultaneously acquired with the red camera (160\,$\mu$m). The data were processed using the Herschel interactive processing environment (HIPE, version 10.0.667), using standard scripts that include bad pixel flagging, detection and correction for glitches and HighPass filtering in order to remove the
$1/f$ noise (see \citealt{Poglitsch2010} for more
details). Figure\,\ref{primary_images} shows the \textsc{Pacs} images at 70 and
160\,$\mu$m. One can immediately notice the two close-by sources that are
background galaxies. The primary, TYC~5241-986-1\,A, is heavily contaminated
by their emission at 160\,$\mu$m, rendering any flux extraction impossible. We
consequently only used the 70\,$\mu$m observation in our analysis. Given the
proximity of the background galaxy to our source of interest we used the
\texttt{IDL} package
\texttt{Starfinder}\footnote{http://www.bo.astro.it/StarFinder/paper6.htm,
  \cite{Diolaiti2000}} to perform PSF photometry of the source. An empirical
PSF of the Vesta asteroid, obtained on OD\,160, with the same scan speed as
our observations (20$\arcsec$\,s$^{-1}$), is used within \texttt{Starfinder}. From
this we detected a source at 22$^h$59${'}$35.23${\arcsec}$ $-$07$^{\circ}$02${'}$24.09${\arcsec}$~(J2000) with a flux of 7.1~mJy
in the 70\,$\mu$m map. This is 2.6 arcseconds from the predicted 2013.0
position of TYC~5241-986-1~A, comparable with the approximate Herschel positional error
of 2.5 arcseconds. To
estimate the noise level and thus the quality of the detection, we randomly
placed apertures on the 70\,$\mu$m map. Due to a smaller detector coverage,
map borders are noisier and not representative of the noise close to the
source. Therefore the different apertures were placed  within a distance of 80
pixels to the central pixel.  We then extracted the aperture photometry for
all of these apertures and fitted a Gaussian profile to the flux distribution
histogram to estimate the $\sigma$ uncertainty of 1.8~mJy for our source. Overall, the estimated uncertainty is consistent with predictions from the \texttt{HerschelSpot}\footnote{http://herschel.esac.esa.int/Tools.shtml version 6.2.0} software.
\subsection{Is the excess caused by a debris disc?}
\label{disc_disc}
   \begin{figure*}
 \setlength{\unitlength}{1mm}
 \begin{picture}(100,120)
 \includegraphics{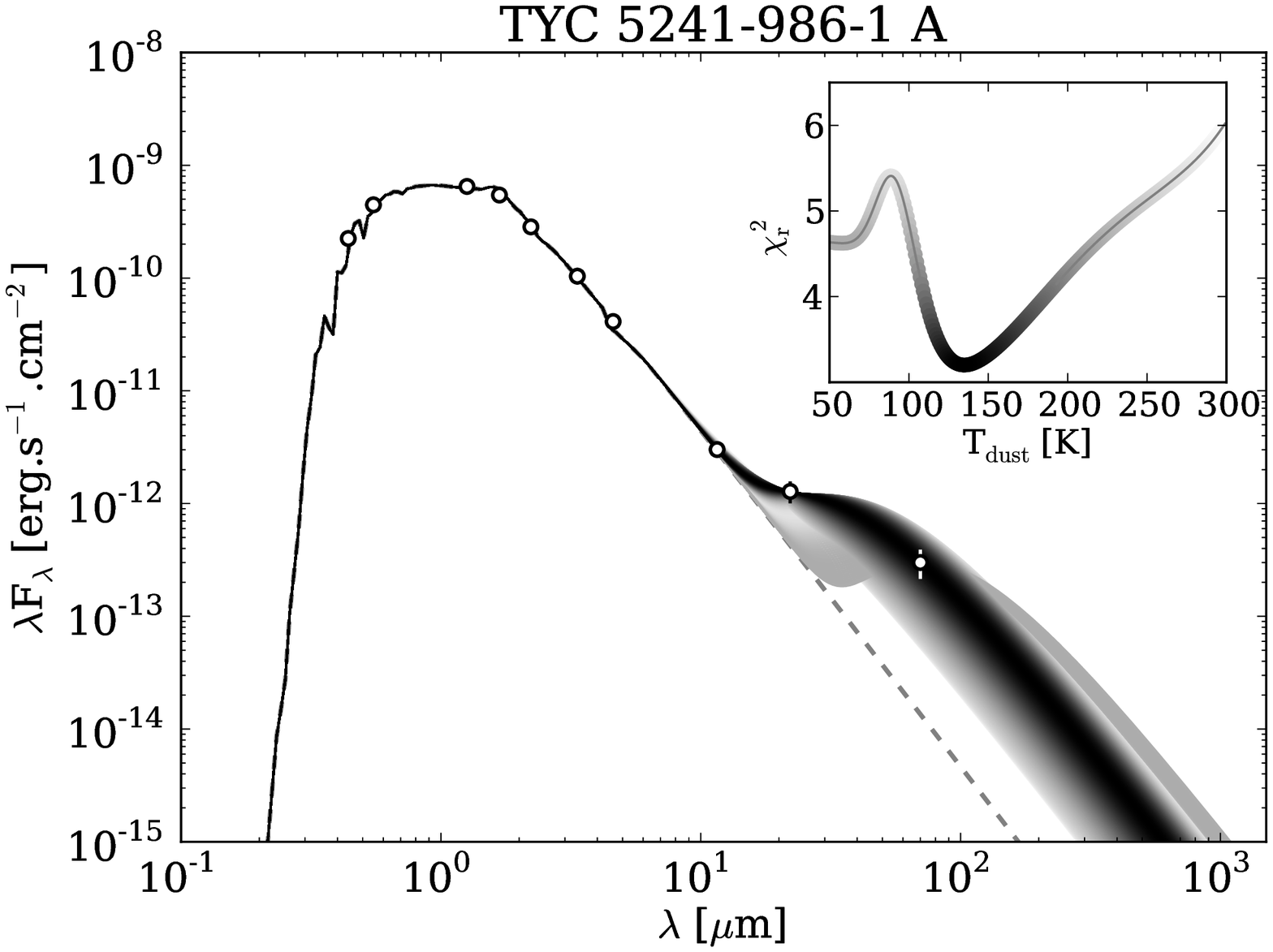}
 \end{picture}
 \caption[]{The SED of TYC~5241-986-1~A. The points are from left to right, Tycho $B$ and $V$ \citep{Hog2000}, 2MASS $J$, $H$ and $K_s$ \citep{Skrutskie2006},WISE $W1$, $W2$, $W3$ and $W4$ \citep{Wright2010,WISE2012} and our Herschel/PACS 70$\mu$m detection. The solid line is a fit to a 4200~K model of the photosphere from \cite{Castelli2004}. The plotted error bars represent 1$\sigma$ confidence limits. A series of Planck functions with different temperatures were fitted to the excess and their reduced $\chi^2$ goodness of fit statistic computed (see inset). The weight of the black and grey lines beyond 10~$\mu$m represents their goodness of fit (darker more likely). Note there are a number of fits with temperatures around 50~K with reasonable $\chi^2$ values. However these do not match the $W4$ excess well. From this plot we characterise the emission as being caused by dust at approximately 100--175~K.}
 \label{disc_sed}
 \end{figure*}
Mid-infrared excesses around stars older than $\sim$10~Myr are characteristic
of debris discs. Figure~\ref{disc_sed} shows that there is a clear excess
above the photosphere at $W4$ and 70~$\mu$m. To characterise the SED, we
fitted a series of Planck functions to the excess and computed the reduced
$\chi^2$ statistic for each. As shown in Figure~\ref{disc_sed}, the excess
fits a temperature in the 100--175~K range with a best fit at 135~K. This
equates to a typical distance of approximately 1.5~AU from this type of
star (Equation~3 from \citealt{Wyatt2008}). \cite{ps_smith2006} identified candidate debris discs around
HD~95650~(M2) and BD~+21$^{\circ}$~2486~(K5) both of which have excess at
70~$\mu$m but not at 24~$\mu$m. The excess in our source exists both at
22~$\mu$m and 70~$\mu$m, indicating that the candidate disc is more like the
one proposed around the early~M dwarf TWA~7, which has excess
at 24$\mu$m and 70$\mu$m. \cite{Low2005} attribute this to an 80~K disc. Given
TYC~5241-986-1~A is likely to be substantially older than the TW~Hydrae
association, it is likely that its disc has recently been replenished by some
process.
From these fits we estimate the dust mass to be $2.5 \times 10^{-5} M_{\oplus}$ using Equation~5 from \cite{Wyatt2008}.
\subsection{Is the excess caused by a background galaxy?}
\label{disc_galaxy}
Caution must always be exercised when associating far-infrared detections with
optical counterparts. This is starkly demonstrated by the case of
TWA~13A. This was proposed to have a debris disc by \cite{Low2005} based on a
single-band excess at 70$\mu$m. However further examination by
\cite{Plavchan2009} identified that the excess was caused by a background
source (likely a galaxy) approximately 10 arcseconds away. Our source differs
from this case in two respects. Firstly we have an excess in both the WISE
$W4$ band and the Herschel 70$\mu$m band. Secondly, our Herchel detection is
within 3 arcseconds of the expected position of the star, significantly closer
than TWA~13A is to its 70$\mu$m counterpart. In-fact our object more closely resembles TWA~7.

\cite{Sibthorpe2013} use galaxy number counts from Herschel surveys to estimate the probability of a coincident background galaxy being associated with a source. To quantify the possibility of a chance alignment, we applied their formula to derive a probability of 0.048\% that a source of 7~mJy would fall within 3 arcseconds of a particular position. It should be noted that our survey returned 38 candidate young stars with proper motions within 5$\sigma$ of their proposed primaries. While only our star and TYC 7443-1102-1~B are obviously good candidates to be real companions, we multiply our 0.048\% by 38 to yield a coincident probability of 1.8\% for our study as a whole. As most of the 38 objects are clear background stars, this probability is likely to be artificially inflated. However it should be noted that \cite{Sibthorpe2013}'s analysis assumes a random distribution of galaxies. As seen in Figure~\ref{primary_images} there are two galaxies within an arcminute of TYC~5241-986-1~A. One of these, 
SDSS~J225934.58-070231.5 has a measured SDSS photometric redshift of 0.25$\pm$0.09. The possible bias caused by galaxy clustering may increase the probability of a chance alignment.

We modelled the SED of this object as a combination of a late~K~dwarf and a
100-175~K Planck function. To first order, a background extragalactic source
must also match a 100--175~K Planck function. While the SED of a galaxy is complex, it is dominated at far infrared wavelengths by emission from cold dust. \cite{dan_smith2012} studied Herschel selected galaxies in the local universe and found that the mean grey body temperature (which describes the dust) was 26.1$\pm$3.5~K. This is substancially different from our best-fit temperature of 100--175~K. This study quotes values for rest-frame wavelength SEDs so clearly redshift could alter the SED of the hypothetical background galaxy. However redshift would gradually push the peak of the dust SED redward, making it an even poorer fit to a 100--175~K blackbody. It is possible that at redshifts of $>$2 the dust component for a typical galaxy will move beyond the Herschel 70$\mu$m band. However this would make it unassociated with the low-redshift galaxies in close proximity to 
the star. In this case the probability of chance alignment would be that
calculated for a random distribution of background galaxies. Our extremely
conservative deterimination of a chance alignment to a background galaxy was
previously calculated to be 1.8\%. In summary, while we cannot conclusively rule
out a chance alignment with a background source, we determine this possibility to be
highly unlikely.
\section{Conclusions}
We have identified a young hierachical triple system. Spectroscopic
observations of the primary indicate that it is an active, moderately young
(20 to $\sim$125~Myr) late-K dwarf. From its radial velocity measurements, proper
motions and photometric distance we associate the system with the Local
Association, but cannot link it to any specific young moving
group. Observations of the secondary and tertiary components indicate that these
are a pair of active M~4.5$\pm$0.5 dwarfs. From these spectral types we derived
the total mass of the system and found it to be extremely loosely bound
compared to field binaries. The primary has excess at WISE
24~$\mu$m and Herschel/PACS 70~$\mu$m. While we cannot completely rule out
contamination by a background galaxy, we find it unlikely based on the SED of
the excess and the likelihood of a chance alignment. We consider a debris disc
with a temperatures between 100--175~K to be the most likely cause of the
excess. Such debris discs around a stars of this apparent age and spectral
type are extremely rare and this source is one of only a handful of candidate
debris discs around late type stars older than $\sim$20~Myr.

\section*{Acknowledgments}We thank Sebastian L\'{e}pine for making his proper
motion catalogue available for the production of the input catalogue. We also
thank Tom Herbst for discussions on our paper draft, Wolfgang Brandner and
Reinhard Mundt for their expertise on activity signatures in M dwarfs, Thomas
Robitaille for helpful discussions, Chad Bender for the use of his spectal
cross-correlation software and Elisabete Da Cunha for her mastery of galaxy SEDs. Thanks to Bruce Sibthorpe for making his background confusion script available. This publication makes use of data from MPIA director's discrectionary time and we thank the staff at the MPG/ESO 2.2~m telescope in La Silla for undertaking our observations. We thank Calar Alto Observatory for allocation of director's discretionary time to this programme. This publication makes use of data products from the Wide-field Infrared Survey Explorer, which is a joint project of the University of California, Los Angeles, and the Jet Propulsion Laboratory/California Institute of Technology, funded by the National Aeronautics and 
Space Administration. We would especially like to thank Roc Cutri and the WISE helpdesk team for assisting us with questions about WISE data. This research has made use of data obtained from the SuperCOSMOS Science Archive, prepared and hosted by the Wide Field Astronomy Unit, Institute for Astronomy, University of Edinburgh, which is funded by the UK Science and Technology Facilities Council. Herschel is an ESA space observatory with science instruments provided by European-led Principal Investigator consortia and with important participation from NASA. The Herschel spacecraft was designed, built, tested, and launched under a contract to ESA managed by the Herschel/Planck Project team by an industrial consortium under the overall responsibility of the prime contractor Thales Alenia Space (Cannes), and including Astrium (Friedrichshafen) responsible for the payload module and for system testing at spacecraft level, Thales Alenia Space (Turin) responsible for the service module, and Astrium (Toulouse) 
responsible for the telescope, with in excess of a hundred subcontractors. This publication makes use of data products from the Two Micron All Sky Survey, which is a joint project of the University of Massachusetts and the Infrared Processing and Analysis Center/California Institute of Technology, funded by the National Aeronautics and Space Administration and the National Science Foundation. Funding for the SDSS and SDSS-II has been provided by the Alfred P. Sloan Foundation, the Participating Institutions, the National Science Foundation, the U.S. Department of Energy, the National Aeronautics and Space Administration, the Japanese Monbukagakusho, the Max Planck Society, and the Higher Education Funding Council for England. The SDSS Web Site is http://www.sdss.org/. The SDSS is managed by the Astrophysical Research Consortium for the Participating Institutions. The Participating Institutions are the American Museum of Natural History, Astrophysical Institute Potsdam, University of Basel, University of 
Cambridge, Case Western Reserve University, University of Chicago, Drexel University, Fermilab, the Institute for Advanced Study, the Japan Participation Group, Johns Hopkins University, the Joint Institute for Nuclear Astrophysics, the Kavli Institute for Particle Astrophysics and Cosmology, the Korean Scientist Group, the Chinese Academy of Sciences (LAMOST), Los Alamos National Laboratory, the Max-Planck-Institute for Astronomy (MPIA), the Max-Planck-Institute for Astrophysics (MPA), New Mexico State University, Ohio State University, University of Pittsburgh, University of Portsmouth, Princeton University, the United States Naval Observatory, and the University of Washington.

\bibliography{ndeacon}

\bibliographystyle{mn2e}
\label{lastpage}

\end{document}